\documentclass[aps,twocolumn,showpacs]{revtex4}
\usepackage{amsmath}
\usepackage{epsfig}

\begin{document}

\title{Reanalysis of the most strange dibaryon within constituent quark models}

\author{Hongxia Huang$^1$}\email{hxhuang@njnu.edu.cn}
\author{Xinmei Zhu$^2$}\email{zxm_yz@126.com}
\author{Jialun Ping$^1$}\email{jlping@njnu.edu.cn(Corresponding author)}

\affiliation{$^1$Department of Physics and Jiangsu Key Laboratory for Numerical
Simulation of Large Scale Complex Systems, Nanjing Normal University, Nanjing 210023, P. R. China}
\affiliation{$^2$Department of Physics, Yangzhou University, Yangzhou 225009, P. R. China}

\begin{abstract}
The most strange dibaryon $\Omega\Omega$ with quantum numbers $S=-6$, $I=0$, and $J^{P}=0^{+},~1^{-},~2^{+},~3^{-}$ is reanalyzed
in the framework of quark delocalization color screening model (QDCSM) and chiral quark model (ChQM). The $\Omega\Omega$ dibaryon
with $J^{P}=0^{+}$ is bound, and the one with other quantum numbers $J^{P}=1^{-},~2^{+},~3^{-}$ are all unbound in our calculation.
The low-energy scattering phase shifts, the scattering length, and the effective range of the $\Omega\Omega$ dibaryon with
$J^{P}=0^{+}$ also support the existence of such strange dibaryon. This dibaryon is showed to be a shallow bound state in QDCSM, 
while the binding energy becomes much larger in the ChQM by including the effect of the hidden-color channel coupling. And the 
scalar nonet meson-exchange in the ChQM also provides more attraction for the $\Omega\Omega$ system. Experimental search for 
such most strange dibaryon will provide much information for understanding the hadron-hadron interactions in different quark models.
\end{abstract}


\maketitle

\setcounter{totalnumber}{5}

\section{\label{sec:introduction}Introduction}
As is commonly believed that quantum chromodynamics (QCD) is the fundamental theory of the strong interaction.
However, the low energy physics of QCD, such as hadron structure, hadron-hadron interactions, and the structure
of multiquark systems, is much harder to calculate directly from QCD. Various QCD-inspired quark models have
been developed to get physical insights into the multiquark systems. There are the MIT bag model~\cite{MIT},
cloudy bag model~\cite{cloudy}, Friedberg-Lee non-topological soliton model~\cite{soliton}, Skyrme topological
soliton model~\cite{skyrme}, the constituent quark model~\cite{Rujula,Isgur}, etc. Different models use quite
different effective degrees of freedom, which might be indicative of the nature of low-energy QCD.

The constituent quark model has been quite successful in understanding hadron spectroscopy and hadron-hadron
interactions even though we have not yet derived the constituent quark model directly from QCD. De Rujula, Georgi,
and Glashow~\cite{Rujula} first put forward a quark-gluon coupling model based on constituent quark and gluon effective
degrees of freedom. Isgur and Karl obtained a good description of hadron spectroscopy based on this model~\cite{Isgur}.
However, extension of the model to baryon-baryon interactions does not reproduce the nucleon-nucleon ($NN$) intermediate
and long-range interaction.

One modification is the addition of scalar meson exchange and Goldstone bosons exchange on the quark
level~\cite{Faessler,Obukhovsky,Fernandez,Fujiwara}, which provide the nucleon-nucleon intermediate and long-range
interaction, respectively. A typical approach is the chiral quark model (ChQM)~\cite{ChQM,Salamanca}, in which
the constituent quarks interact with each other through colorless Goldstone bosons exchange in addition to the
colorful one-gluon-exchange and confinement. To obtain the immediate-range attraction of $NN$ interaction,
the chiral partner $\sigma$ meson-exchange has to be introduced. The $\sigma$ meson had been observed by BES
collaboration as a $\pi\pi$ $S$-wave resonance~\cite{BES}. However, the results found by three groups independently,
show that the correlated two-pion exchange between two nucleons generates strong short-range repulsion and very
moderate long-range attraction, which is quite different from the behavior of the $\sigma$ meson which used in
the ChQM~\cite{sigma}. Therefore, one may wonder that the $\sigma$ meson used in the ChQM is the correlated $\pi\pi$
resonance or an effective one.

An alternative approach to study baryon-baryon interaction is the quark delocalization color
screening model (QDCSM), which was developed in 1990s with the aim of explaining the similarities
between nuclear and molecular forces~\cite{QDCSM0}. Two new ingredients were introduced: quark
delocalization (to enlarge the model variational space to take into account the mutual distortion
or the internal excitations of nucleons in the course of their interactions) and color screening
(assuming the quark-quark interaction dependent on quark states aimed to take into account the QCD
effect which has not yet been included in the two-body confinement and effective one gluon exchange).
The model gives a good description of $NN$ and $YN$ interactions and the properties of
deuteron~\cite{QDCSM1,QDCSM2}. It is also employed to calculate the baryon-baryon scattering phase shifts
and predict the dibaryon candidates $d^{*}$ and $N\Omega$~\cite{Ping1,Huang1,PRC51,PangHR,ChenM,Huang2}.

The difference between the ChQM and QDCSM is the intermediate-range attraction mechanism, which is the
$\sigma$ meson-exchange in ChQM and the quark delocalization and color screening in QDCSM. These two models
have been applied to the study of nucleon-nucleon ($NN$) and the $N\Omega$ systems~\cite{QDCSM2,Ping1,ChenM,Huang2}.
The results show that the intermediate-range attraction mechanism in the QDCSM is equivalent to the $\sigma$
meson-exchange in the ChQM in these two systems. And the color screening is an effective description of the
hidden-color channels coupling~\cite{Huang3}. It is interesting to check this consistency in other systems,
such as the most strange dibaryon $\Omega\Omega$.

The dibaryon $\Omega\Omega$ with quantum numbers $S=-6$, $I=0$, and $J=0$ was predicted by Kopeliovich in the
framework of the flavor $SU(3)$ Skyrmion model~\cite{Kopeliovich}. Zhang $et~al.$ also suggested to search for
this $\Omega\Omega$ state in heavy ion collision experiments~\cite{Zhang}. And this dibaryon was also showed to
be a bound state in the QDCSM~\cite{PangHR2}. Very recently, the $\Omega\Omega$ with $S=-6$, $I=0$, and $J=0$ was
investigated by the HAL QCD Collaboration~\cite{Lattic}. They studied this dibaryon on the basis of the ($2+1$)-flavor
lattice QCD simulations with a nearly physical pion mass $m_{\pi}\simeq 146$ MeV. The results showed that this
$\Omega\Omega$ state had an overall attraction and was located near the unitary regime. They suggested that such
a system can be best searched experimentally by the pair-momentum correlation in relativistic heavy-ion collisions.
Then Morita $et~al.$~\cite{Morita} calculated the correlation functions of this $\Omega\Omega$ state based on an
expanding source model by using the interaction potentials from the lattice QCD calculations.

In this work, we reanalyze the most strange dibaryon $\Omega\Omega$ with quantum numbers $S=-6$, $I=0$, and
$J^{P}=0^{+},~1^{-},~2^{+},~3^{-}$ in both ChQM and QDCSM. The binding energy, as well as the low-energy scattering
phase shifts, the scattering length, and the effective range, which are useful for the experimental search of this
strange dibaryon, are investigated. By comparing the results within these two quark models, we can check the model
dependence of this dibaryon. On the other hand, we can also inspect the consistency of the intermediate-range
attraction mechanism of these two models in such strange dibaryon $\Omega\Omega$ system.

The structure of this paper is as follows. A brief introduction of two quark models is given in section II.
Section III devotes to the numerical results and discussions. The summary is shown in the last section.

\section{Two quark models}

\subsection{Chiral quark  model}

In this work, the Salamanca model was chosen as the representative of the ChQM, because the work of the Salamanca
group covers the hadron spectra and the nucleon-nucleon interaction, and has been extended to the study
of multiquark states. The model details can be found in Ref. ~\cite{Salamanca}. Here only the Hamiltonian is given:

\begin{widetext}
\begin{eqnarray}
H & = & \sum_{i=1}^6\left(m_i+\frac{p_i^2}{2m_i}\right)-T_{CM} +\sum_{j>i=1}^6
\left(V^{C}_{ij}+V^{G}_{ij}+V^{\chi}_{ij}+V^{\sigma}_{ij}\right), \\
V^{C}_{ij} & = & -a_{c} \boldsymbol{\lambda}^c_{i}\cdot \boldsymbol{
\lambda}^c_{j} ({r^2_{ij}}+v_{0}), \label{sala-vc} \\
V^{G}_{ij} & = & \frac{1}{4}\alpha_s \boldsymbol{\lambda}^{c}_i \cdot
\boldsymbol{\lambda}^{c}_j
\left[\frac{1}{r_{ij}}-\frac{\pi}{2}\delta(\boldsymbol{r}_{ij})(\frac{1}{m^2_i}+\frac{1}{m^2_j}
+\frac{4\boldsymbol{\sigma}_i\cdot\boldsymbol{\sigma}_j}{3m_im_j})-\frac{3}{4m_im_jr^3_{ij}}
S_{ij}\right] \label{sala-vG} \\
V^{\chi}_{ij} & = & V_{\pi}( \boldsymbol{r}_{ij})\sum_{a=1}^3\lambda
_{i}^{a}\cdot \lambda
_{j}^{a}+V_{K}(\boldsymbol{r}_{ij})\sum_{a=4}^7\lambda
_{i}^{a}\cdot \lambda _{j}^{a}
+V_{\eta}(\boldsymbol{r}_{ij})\left[\left(\lambda _{i}^{8}\cdot
\lambda _{j}^{8}\right)\cos\theta_P-(\lambda _{i}^{0}\cdot
\lambda_{j}^{0}) \sin\theta_P\right] \label{sala-Vchi1} \\
V_{\chi}(\boldsymbol{r}_{ij}) & = & {\frac{g_{ch}^{2}}{{4\pi
}}}{\frac{m_{\chi}^{2}}{{\
12m_{i}m_{j}}}}{\frac{\Lambda _{\chi}^{2}}{{\Lambda _{\chi}^{2}-m_{\chi}^{2}}}}%
m_{\chi} \left\{(\boldsymbol{\sigma}_{i}\cdot
\boldsymbol{\sigma}_{j})
\left[ Y(m_{\chi}\,r_{ij})-{\frac{\Lambda_{\chi}^{3}}{m_{\chi}^{3}}}%
Y(\Lambda _{\chi}\,r_{ij})\right] \right.\nonumber \\
&& \left. +\left[H(m_{\chi}
r_{ij})-\frac{\Lambda_{\chi}^3}{m_{\chi}^3}
H(\Lambda_{\chi} r_{ij})\right] S_{ij} \right\}, ~~~~~~\chi=\pi, K, \eta, \\
V^{\sigma}_{ij} & = & -{\frac{g_{ch}^{2}}{{4\pi }}}
{\frac{\Lambda _{\sigma}^{2}}{{\Lambda _{\sigma}^{2}-m_{\sigma}^{2}}}}%
m_{\sigma}\left[ Y(m_{\sigma}\,r_{ij})-{\frac{\Lambda _{\sigma}}{m_{\sigma}}}%
Y(\Lambda _{\sigma}\,r_{ij})\right] , \\
S_{ij}&=&\left\{ 3\frac{(\boldsymbol{\sigma}_i
\cdot\boldsymbol{r}_{ij}) (\boldsymbol{\sigma}_j\cdot
\boldsymbol{r}_{ij})}{r_{ij}^2}-\boldsymbol{\sigma}_i \cdot
\boldsymbol{\sigma}_j\right\},\\
H(x)&=&(1+3/x+3/x^{2})Y(x),~~~~~~
 Y(x) =e^{-x}/x. \label{sala-vchi2}
\end{eqnarray}
\end{widetext}
Where $\alpha_s$ is the quark-gluon coupling constant. In order to cover the wide energy scale from light
to strange quark, one introduces an effective scale-dependent quark-gluon coupling constant $\alpha_s(\mu)$~\cite{JPG31},
\begin{equation}
\alpha_s(\mu)=\frac{\alpha_{0}}{\ln\left(\frac{\mu^{2}+\mu_0^{2}}{\Lambda_0^{2}}\right)},
\label{alpha-s}
\end{equation}
where $\mu$ is the reduced mass of the interacting quark-pair. The coupling constant $g_{ch}$ for chiral field is
determined from the $NN\pi$ coupling constant through
\begin{equation}
\frac{g_{ch}^{2}}{4\pi }=\left( \frac{3}{5}\right) ^{2}{\frac{g_{\pi NN}^{2}%
}{{4\pi }}}{\frac{m_{u,d}^{2}}{m_{N}^{2}}}\label{gch}.
\end{equation}
The other symbols in the above expressions have their usual meanings.

For the most strange dibaryons, two versions of ChQM~\cite{Garcilazo,QBLi} are used here. One is the $SU(2)$ ChQM, in which
$\sigma$ meson is restricted to exchange between $u$ and/or $d$ quark pair only; another is the $SU(3)$ ChQM, where full
$SU(3)$ scalar nonet meson-exchange was used. These scalar potentials have the same functional form as the one of $SU(2)$
ChQM but a different $SU(3)$ operator dependence~\cite{Garcilazo}, that is,
\begin{eqnarray}
V^{\sigma_{a}}_{ij} & = & V_{a_{0}}(
\boldsymbol{r}_{ij})\sum_{a=1}^3\lambda _{i}^{a}\cdot \lambda_{j}^{a}+V_{\kappa}(\boldsymbol{r}_{ij})
\sum_{a=4}^7\lambda_{i}^{a}\cdot \lambda _{j}^{a} \nonumber \\
& & +V_{f_{0}}(\boldsymbol{r}_{ij})\lambda _{i}^{8}\cdot \lambda_{j}^{8}+V_{\sigma}(\boldsymbol{r}_{ij})
\lambda _{i}^{0}\cdot \lambda _{j}^{0} \label{sala-su3} \\
V_{k}(\boldsymbol{r}_{ij}) & = & -{\frac{g_{ch}^{2}}{{4\pi }}}
\frac{\Lambda _{k}^{2}m_{k}}{{\Lambda_{k}^{2}-m_{k}^{2}}}
\left[ Y(m_{k}\,r_{ij})-{\frac{\Lambda _{k}}{m_{k}}}%
Y(\Lambda _{k}\,r_{ij})\right] , \nonumber
\end{eqnarray}
with $k=a_{0}, \kappa, f_{0}$ or $\sigma$.

\subsection{Quark delocalization color screening model}
The Hamiltonian of QDCSM is almost the same as that of ChQM but with two modifications~\cite{QDCSM0,QDCSM1}: First, there
is no $\sigma$-meson exchange in QDCSM, and second, the screened color confinement is used between quark pairs resident
in different baryon orbits. That is
\begin{equation}
V_{ij}^{C}=\left \{ \begin{array}{ll}
-a_{c}\boldsymbol{\mathbf{\lambda}}^c_{i}\cdot
\boldsymbol{\mathbf{ \lambda}}^c_{j}~(r_{ij}^2+ v_0) &
  \mbox{if \textit{i},\textit{j} in the same} \\
  &  \mbox{baryon orbit} \\
-a_{c}\boldsymbol{\mathbf{\lambda}}^c_{i}\cdot
\boldsymbol{\mathbf{
\lambda}}^c_{j}~(\frac{1-e^{-\mu_{ij}\mathbf{r}_{ij}^2}}{\mu_{ij}}+
v_0) & \mbox{otherwise} \end{array} \right.\label{QDCSM-vc}
\end{equation}
where the color screening constant $\mu_{ij}$ is determined by fitting the deuteron properties, $NN$ scattering
phase shifts and $N\Lambda$, $N\Sigma$ scattering cross sections, $\mu_{uu}=0.45$, $\mu_{us}=0.19$ and
$\mu_{ss}=0.08$, which satisfy the relation, $\mu_{us}^{2}=\mu_{uu}\mu_{ss}$.

The single particle orbital wave functions in the ordinary quark cluster model are the left and right centered single
Gaussian functions:
\begin{eqnarray}
\phi_\alpha(\boldsymbol {S_{i}})=\left(\frac{1}{\pi
b^2}\right)^{\frac{3}{4}}e^ {-\frac{(\boldsymbol {r}-\boldsymbol
{S_i}/2)^2}{2b^2}},
 \nonumber\\
\phi_\beta(-\boldsymbol {S_{i}})=\left(\frac{1}{\pi
b^2}\right)^{\frac{3}{4}}e^ {-\frac{(\boldsymbol {r}+\boldsymbol
{S_i}/2)^2}{2b^2}} .
 \
\end{eqnarray}
The quark delocalization in QDCSM is realized by writing the single particle orbital wave function as a
linear combination of the left and right Gaussians:
\begin{eqnarray}
{\psi}_{\alpha}(\boldsymbol {S_{i}},\epsilon) &=&
\left({\phi}_{\alpha}(\boldsymbol{S_{i}})
+\epsilon{\phi}_{\alpha}(-\boldsymbol{S_{i}})\right)/N(\epsilon),
\nonumber \\
{\psi}_{\beta}(-\boldsymbol {S_{i}},\epsilon) &=&
\left({\phi}_{\beta}(-\boldsymbol{S_{i}})
+\epsilon{\phi}_{\beta}(\boldsymbol{S_{i}})\right)/N(\epsilon),
\nonumber \\
N(\epsilon)&=&\sqrt{1+\epsilon^2+2\epsilon e^{{-S}_i^2/4b^2}}.
\end{eqnarray}
where $\epsilon(\boldsymbol{S}_i)$ is the delocalization parameter determined by the dynamics of the quark system rather than
adjusted parameters. In this way, the system can choose its most favorable configuration through its own dynamics in a larger
Hilbert space.

The parameters of these models are from our previous work of $N\Omega$ system~\cite{Huang2}. We list all parameters in
Table~\ref{parameters}. The calculated baryon masses in comparison with experimental values are shown in Table~\ref{mass}.
\begin{table}[ht]
\caption{\label{parameters} The parameters of two models:
$m_{\pi}=0.7$fm$^{-1}$, $m_{K}=2.51$fm$^{-1}$,
$m_{\eta}=2.77$fm$^{-1}$, $m_{\sigma}=3.42$fm$^{-1}$,
$m_{a_{0}}=m_{\kappa}=m_{f_{0}}=4.97$fm$^{-1}$,
$\Lambda_{\pi}=4.2$fm$^{-1}$, $\Lambda_{K}=5.2$fm$^{-1}$,
$\Lambda_{\eta}=5.2$fm$^{-1}$, $\Lambda_{\sigma}=4.2$fm$^{-1}$,
$\Lambda_{a_{0}}=\Lambda_{\kappa}=\Lambda_{f_{0}}=5.2$fm$^{-1}$,
$g_{ch}^2/(4\pi)$=0.54, $\theta_p$=$-15^{0}$. }
\begin{tabular}{lcccc}
\hline
              &                         &~~QDCSM~~       &~~$SU(2)$ ChQM~~  &~~$SU(3)$ ChQM~~  \\   \hline
              &$b$ (fm)                  &~~0.518     &~~0.518       &~~0.518    \\
              & $m_u$ (MeV)              &~~313       &~~313         &~~313      \\
              & $m_d$ (MeV)              &~~313       &~~313         &~~313      \\
              & $m_s$ (MeV)              &~~573       &~~536         &~~573      \\
              &$a_c$ (MeV)               &~~58.03     &~~48.59       &~~48.59    \\
              &$\mu_{uu}$ (fm$^{-2}$)    &~~0.45      &~~-           &~~-        \\
              &$\mu_{us}$ (fm$^{-2}$)    &~~0.19      &~~-           &~~-        \\
              &$\mu_{ss}$ (fm$^{-2}$)    &~~0.08      &~~-           &~~-        \\
              &$v_{0}$ (MeV)             &-1.2883     &-1.2145       &-0.961\\
              &$\alpha_0$               &~~0.510     &~~0.510      &~~0.583     \\
              &$\Lambda_0$ (fm$^{-1}$)   &~~1.525     &~~1.525      &~~1.616     \\
              &$\mu_0$ (MeV)             &~~445.808   &445.808      &422.430     \\
                                                                       \hline

\hline
\end{tabular}
\end{table}

\begin{table}[ht]
\caption{\label{mass}The masses of the ground-state baryons(in MeV).}
\begin{tabular}{lcccccccc}
\hline
               & ~~$N$~~            & ~~$\Delta$~~ & ~~$\Lambda$~~ & ~~$\Sigma$~~
               & ~~$\Sigma^*$~~     & ~~$\Xi$~~    & ~~$\Xi^*$~~   & ~~$\Omega$~~  \\ \hline
QDCSM          & 939          & 1232       & 1124    & 1238
                & 1360         & 1374       & 1496   & 1642       \\
$SU(2)$ ChQM      & 939          & 1232       & 1137     & 1245
                & 1376         & 1375       & 1506     & 1620     \\
$SU(3)$ ChQM      & 939          & 1232       & 1123     & 1267
                & 1344         & 1398       & 1475     & 1625     \\
 Expt.  & 939  &1232 &1116 &1193 &1385 &1318 &1533 &1672\\  \hline
\end{tabular}
\end{table}

\section{The results and discussions}
\subsection{Effective potentials}
\begin{figure*}
\epsfxsize=4.5in \epsfbox{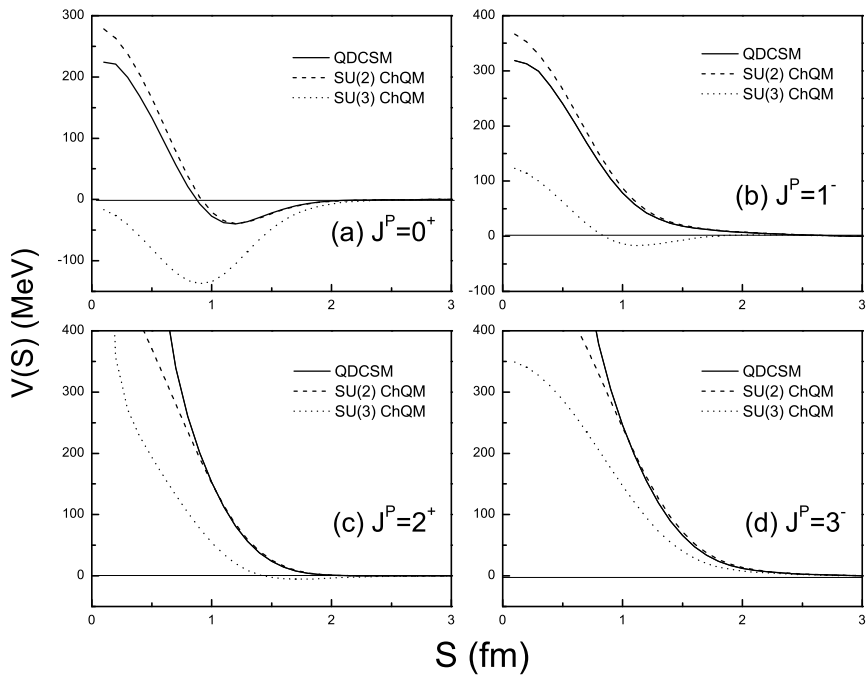} \vspace{-0.1in}
\caption{The effective potentials of the $\Omega\Omega$ system in different quark models.}
\end{figure*}

In this work, we investigate the most strange dibaryon $\Omega\Omega$ with quantum numbers $S=-6$, $I=0$, and
$J^{P}=0^{+},~1^{-},~2^{+},~3^{-}$ in QDCSM, $SU(2)$ ChQM and $SU(3)$ ChQM. The partial wave of $J^{P}=0^{+},~2^{+}$
is $S$-wave, and the one of $J^{P}=1^{-},~3^{-}$ is $P$-wave. We calculate the effective potentials of the $\Omega\Omega$
system, because an attractive potential is necessary for forming bound state or resonance. The effective potential between
two clusters is defined as, $V(S)=E(S)-E(\infty)$, where $E(S)$ is the diagonal matrix element of the Hamiltonian of the
system in the generating coordinate. The effective potentials of $J^{P}=0^{+},~1^{-},~2^{+},~3^{-}$ in three quark models
are shown in Fig. 1(a), (b), (c) and (d), respectively.

From Fig. 1(a), we can see that the potentials are attractive for the $J^{P}=0^{+}$ $\Omega\Omega$ state in all quark models.
It is obvious that the attraction in $SU(3)$ ChQM is the largest one, followed by the attractions in $SU(2)$ ChQM and QDCSM.
For the $\Omega\Omega$ state with $J^{P}=1^{-}$ , the potential is repulsive in $SU(2)$ ChQM and QDCSM, but in $SU(3)$ ChQM,
it is a little bit attractive. The case is similar for the $\Omega\Omega$ state with $J^{P}=2^{+}$. For the $\Omega\Omega$
state with $J^{P}=3^{-}$, the potentials are all repulsive in three quark models. Therefore, from the behavior of the effective
potentials of the $\Omega\Omega$ state, it is possible for the $J^{P}=0^{+}$ $\Omega\Omega$ dibaryon to form bound state, while
for the $\Omega\Omega$ dibaryon with other quantum numbers, it is nearly impossible to form any bound state because of the
repulsive interaction between two $\Omega$s.

\begin{figure*}
\epsfxsize=4.8in \epsfbox{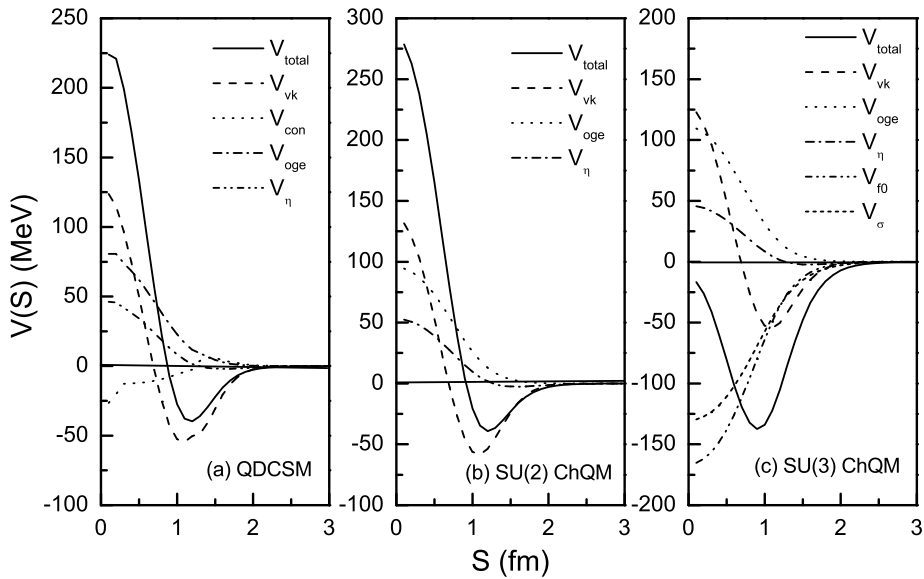} \vspace{-0.1in}
\caption{The contributions to the effective potential from various
terms of interactions for the $J^{P}=0^{+}$ $\Omega\Omega$ state.}
\end{figure*}

In order to investigate the source of the attractions for the $J^{P}=0^{+}$ $\Omega\Omega$ state, we calculate the contribution
of each interaction term to the total potential of the system. The potentials of various terms: the kinetic energy
($V_{vk}$), the confinement ($V_{con}$), the one-gluon-exchange ($V_{oge}$), the one-boson-exchange ($V_{\pi}$ and $V_{K}$ do not
contribute to the effective potential because they do not exchange between two $s$ quarks, so only $V_{\eta}$ contributes),
and the scalar nonet meson-exchange ($V_{\sigma}$, $V_{a_{0}}$, $V_{\kappa}$, and $V_{f_{0}}$) are shown in Fig. 2.

For the QDCSM, quark delocalization and color screening work together to provide short-range repulsion and intermediate-range
attraction. We illustrate this mechanism by showing contributions of all interaction terms to the effective potential
in Fig. 2(a). From which we see that the attraction of the $\Omega\Omega$ system mainly comes from the kinetic energy term.
The confinement interaction provides a little attraction, while other terms provide repulsive potentials, which reduce
the total attraction of the $\Omega\Omega$ potential.

For the $SU(2)$ ChQM, the quadratic confinement do not contribute to the potential between two $\Omega$'s, because of the
properties of two color singlets. Since the $\sigma$ meson is restricted to exchange between the $u$ and $d$ quarks only in
$SU(2)$ ChQM, there is no $\sigma$ meson-exchange interaction between two $\Omega$s. Therefore, there are only kinetic energy,
one-gluon-exchange and one-$\eta$-exchange contribute to the effective potentials. It is shown in Fig. 2(b) that the kinetic
energy term provides the major attraction, while other two terms provide repulsive potentials, which decrease the total attractions.

For the $SU(3)$ ChQM, the scalar nonet meson-exchange is included. Although $a_{0}$ and $\kappa$ mesons do not contribute because
they do not exchange between $s$ quarks, both the $f_{0}$ meson-exchange and the $\sigma$ meson-exchange introduce
large attractions, which lead to the strong attraction between two $\Omega$s.

\subsection{Binding energy calculation}
In order to see whether or not there is any bound state, we carry out a dynamic calculation. The resonating group method
(RGM), described in more detail in Ref.~\cite{RGM}, is used here. Expanding the relative motion wavefunction between two
clusters in the RGM by a set of gaussians, the integro-differential equation of RGM can be reduced to algebraic equation,
the generalized eigen-equation. The energy of the system can be obtained by solving the eigen-equation. In the calculation,
the baryon-baryon separation ($|\mathbf{s}_n|$) is taken to be less than 6 fm (to keep the matrix dimension manageably small).

In our calculation, the $\Omega\Omega$ system with $J^{P}=1^{-},2^{+}$,$3^{-}$ are unbound in all quark models, which agree
with the repulsive nature of the interaction of these states. While, the $J^{P}=0^{+}$ $\Omega\Omega$ state is bound in all
quark models, due to the strong attractions in this system. Here, we discuss the $J^{P}=0^{+}$ state in detail.
The binding energies of $J^{P}=0^{+}$ $\Omega\Omega$ state in various quark models are listed in Table~\ref{bound}, where $B_{sc}$
stands for the binding energy of the single channel $\Omega\Omega$, and $B_{cc}$ refers
to the binding energy with the hidden-color channel coupling.
\begin{table}[ht]
\caption{The binding energies (in MeV) of the $J^{P}=0^{+}$ $\Omega\Omega$ system in different models.}
\begin{tabular}{lcccc}
\hline
            & ~~~$B_{sc}$~~~ & ~~~$B_{cc}$~~~   \\ \hline
 ~~~~QDCSM~~~~      & $-2.0$  & $\cdots$   \\
 ~~~~SU(2) ChQM~~~~  & $-0.7$  & $-46.8$  \\
 ~~~~SU(3) ChQM~~~~  & $-79.9$ & $-103.3$   \\
  \hline
\end{tabular}
\label{bound}
\end{table}

The single channel calculation shows that the binding energy in QDCSM and $SU(2)$ ChQM is very small,
which indicates that the $J^{P}=0^{+}$ $\Omega\Omega$ is a shallow bound state. In contrast, the binding energy in the $SU(3)$ 
ChQM is much larger due to the stronger attraction between two $\Omega$s, which suggests that the $\Omega\Omega$ 
is a deep bound state. By coupling the hidden-color channel, a much deeper binding energy is obtained in both the $SU(2)$ ChQM 
and $SU(3)$ ChQM, which reaches to $-46.8$ MeV and $-103.3$ MeV respectively. It indicates that the effect of the hidden-color 
channel coupling is important for the $\Omega\Omega$ system in the ChQM. In QDCSM, since it contains hidden-color channels coupling 
effect already through the color screening~\cite{Ping1,Huang3}, including the color-singlet channels is enough. Therefore, 
we find that the $J^{P}=0^{+}$ $\Omega\Omega$ appears as a shallow bound state in QDCSM, and this conclusion is consistent with 
that of the HAL QCD Collaboration~\cite{Lattic}, in which they showed that the $J^{P}=0^{+}$ $\Omega\Omega$ state had an overall 
attraction and was located near the unitary regime. However, the $J^{P}=0^{+}$ $\Omega\Omega$ state becomes a deeper bound state 
in the $SU(2)$ ChQM, and even a much deeper bound state in the $SU(3)$ ChQM.

\subsection{The low-energy scattering phase shifts}
For the purpose of providing more information for the experimental search of such most strange dibaryon, we calculate the low-energy
scattering phase shifts, the scattering length, and the effective range of the $\Omega\Omega$ dibaryon with $J^{P}=0^{+}$.
In experiment, each $\Omega$ can be identified through a successive weak decay $\Omega^{-}\rightarrow \Lambda + K^{-}
\rightarrow p+\pi^{-}+K^{-}$. A large scattering length (not the existence of a bound state) is the important element for
the correlation $C(Q)$ to have characteristic enhancement at small relative momentum $Q$~\cite{Morita2}.
Here, the well developed Kohn-Hulthen-Kato (KHK) variational method is used to calculate the low-energy scattering phase shifts.
The details can be found in Ref.~\cite{RGM}. Fig. 3 illustrates the scattering phase shifts of the $J^{P}=0^{+}$ $\Omega\Omega$
state. It is obvious that in all quark models, the scattering phase shifts go to $180$ degrees at $E_{c.m.}\sim 0$ and rapidly
decreases as $E_{c.m.}$ increases, which implies the existence of a bound state. The results are
consistent with the the bound state calculation shown above. Besides, the behavior of the low-energy scattering phase shifts is
also in agreement with that of the lattice QCD calculation~\cite{Lattic}.

\begin{figure}
\epsfxsize=3.0in \epsfbox{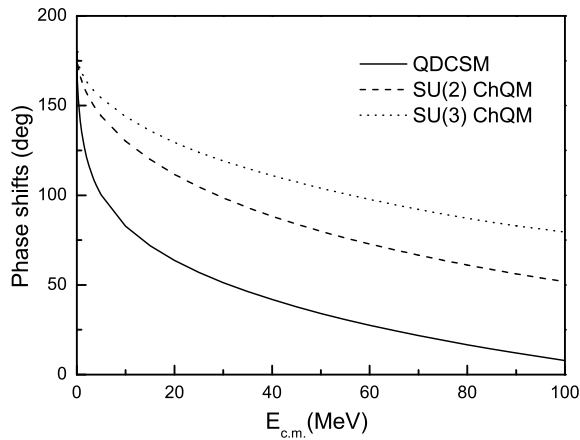} \vspace{-0.1in}
\caption{The phase shifts of the $J^{P}=0^{+}$ $\Omega\Omega$ state.}
\end{figure}

Then, the scattering length $a_{0}$ and the effective range $r_{0}$ of the $\Omega\Omega$ state can be extracted from
the low-energy scattering phase shifts by the following formula:
\begin{eqnarray}
kcot\delta & = & -\frac{1}{a_{0}}+\frac{1}{2}r_{0}k^{2}+{\cal
O}(k^{4})
\end{eqnarray}
where $\delta$ is the low-energy scattering phase shifts, $k$ is the momentum of the relative motion with
$k=\sqrt{2\mu E_{cm}}$, $\mu$ is the reduced mass of two baryons, and $E_{cm}$ is the incident energy.
The binding energy $B^{\prime}$ can be calculated according to the relation:
\begin{eqnarray}
B^{\prime} & = &\frac{\hbar^2\alpha^2}{2\mu} \label{bprime}
\end{eqnarray}
where $\alpha$ is the wave number, which can be obtained from the relation~\cite{Babenko}:
\begin{eqnarray}
r_{0} & = &\frac{2}{\alpha}(1-\frac{1}{\alpha a_{0}})
\end{eqnarray}
The results are listed in Table~\ref{length}.
\begin{center}
\begin{table}[h]
\caption{The scattering length $a_{0}$, effective range $r_{0}$,
and binding energy $B^{\prime}$ of the $J^{P}=0^{+}$ $\Omega\Omega$ dibaryon.}
\begin{tabular}{lcccc}
\hline
  & ~~~$a_{0}~(fm)$~~~ & ~~~$r_{0}~(fm)$~~~ & ~~~$B^{\prime}~(MeV)$~~~    \\ \hline
 QDCSM     & 4.1296  & 1.3977 & $-0.31$    \\
 SU(2) ChQM & 1.4050  & 0.6813 & $-35.6$   \\
 SU(3) ChQM & 0.87275  & 0.42850 & $-92.3$   \\
  \hline
\end{tabular}
\label{length}
\end{table}
\end{center}

From Table~\ref{length}, we can see that in all quark models, the scattering length are all positive,
which implies that the $J^{P}=0^{+}$ $\Omega\Omega$ dibaryon is a bound state here. The binding energies obtained
by Eq.~(\ref{bprime}) is broadly consistent with that in Table~\ref{bound}, which is obtained by the dynamic calculation.
Here again, the scattering length $a_{0}$ and effective range $r_{0}$ of the $J^{P}=0^{+}$ $\Omega\Omega$ dibaryon in
the QDCSM are in agreement with the results of the lattice QCD calculation~\cite{Lattic}, in which
$a_{0}=4.6(6)(^{+1.2}_{-0.5})$ fm and $r_{0}=1.27(3)(^{+0.06}_{-0.03})$ fm.

\section{Summary}
In summary, we reanalyze the most strange dibaryon $\Omega\Omega$ with quantum numbers $S=-6$, $I=0$, and
$J^{P}=0^{+},~1^{-},~2^{+},~3^{-}$ in QDCSM, $SU(2)$ ChQM and $SU(3)$ ChQM. Our results show:
(1) The $\Omega\Omega$ with quantum numbers $J^{P}=1^{-},~2^{+},~3^{-}$ are all unbound, due to the repulsive
interaction between two $\Omega$s in these systems.
(2) For the $J^{P}=0^{+}$ $\Omega\Omega$, the attraction between two $\Omega$s is strong enough to form a bound state
in all quark models.
(3) The low-energy scattering phase shifts, the scattering length, and the effective range of the $\Omega\Omega$ system
with $J^{P}=0^{+}$ also support that this most strange dibaryon is a bound state.
(4) All the results in the QDCSM are consistent with that obtained by the HAL QCD method, which suggest
that the $J^{P}=0^{+}$ $\Omega\Omega$ dibaryon is a shallow bound state. In contract, the binding energy is much deeper in both the $SU(2)$ ChQM
$SU(3)$ ChQM, where the hidden-color channel coupling is employed.

The quark model study of the hadron interaction has experienced a long history. The mechanism of the intermediate-range
attraction of the baryon-baryon interaction is one of the important issues in the study. In $NN$ case we have
shown that the phenomenological $\sigma$ meson exchange in ChQM is equivalent to the quark delocalization and color screening
in QDCSM \cite{QDCSM2} and the color screening effect in QDCSM is an effective description of hidden color channel
coupling~\cite{Huang3}. For the strange $N\Omega$ system, the QDCSM predicts a bound $N\Omega$ dibaryon with quantum numbers
$S=-3$, $I=\frac{1}{2}$, $J^{P}=2^{+}$, while the ChQM cannot obtain the bound state if the $\sigma$-meson is not universally
exchanged between any quark pair. However, the bound state was finally obtained by considering the the hidden-color channels
coupling. Although the similar results are obtained in both models, the mechanism is different. In QDCSM, quark delocalization
and color screening work together to provide short-range repulsion and intermediate-range attraction, the coupling of the color
singlet channels is enough to form a bound state $N\Omega$, while in the $SU(3)$ ChQM, although the universal $\sigma$-meson
exchange introduces large attraction, but it is canceled by the repulsive potentials of $\kappa$ and $f_{0}$ exchange, and
the bound $N\Omega$ state is obtained by coupling both color singlet and hidden-color channels.
Extending to the most strange $\Omega\Omega$ dibaryon, a shallow bound state is obtained in QDCSM. But this $J^{P}=0^{+}$ 
$\Omega\Omega$ become a much deeper bound state in both the $SU(2)$ ChQM and $SU(3)$ ChQM by coupling the hidden-color channel. 
Besides, in the $SU(3)$ ChQM, the $f_{0}$
meson-exchange and the $\sigma$ meson-exchange introduce large attractions, which also increase the total attraction of the
$\Omega\Omega$ potential.

To validate the intermediate attraction mechanism, more experimental data are needed. Experimental search for dibaryons
may provide more information for this issue. Searching for the $N\Omega$ bound state has made considerable headway by the STAR
experiment~\cite{RHIC}. If experiment confirms the existence of $N\Omega$ dibaryon state, it will be a signal showing that the
quark delocalization and color screening (an effective description of hidden color channels coupling) is effective way to
describe the intermediate range attraction of baryon-baryon interaction. This mechanism is also preferred
by the similarity between nuclear force and molecular force. Besides, from the phenomenological
point of view, the $\Omega\Omega$ system can be best searched by the measurement of pair-momentum correlation $C(Q)$ with $Q$
being the relative momentum between two baryons produced in relativistic heavy-ion collisions~\cite{Cho}.
Experimental confirmation of the $N\Omega$ and $\Omega\Omega$ dibaryons will provide other samples of six-quark system
than the non-strange dibaryon $d^*$ \cite{dstar1,dstar2,dstar3}.
We wish there will be more experimental collaborations to be involved in the search of such strange dibaryons.

\acknowledgments{This work is supported partly by the National Natural Science Foundation of China under
Contract Nos. 11675080, 11775118 and 11535005.}

\end{document}